\documentstyle{l-aa-135}

\input psfig

\def\jref#1 #2 #3 #4 {{\par\noindent \hangindent=3em \hangafter=1
      \advance \rightskip by 0em #1, {\it#2}, {\bf#3}, #4.\par}}
\def\rref#1{{\par\noindent \hangindent=3em \hangafter=1
      \advance \rightskip by 0em #1.\par}}
\newcommand{\ggg}{$\gamma$}

\newcommand{\ergs}{\rm \su  erg \su s^{-1}}

\newcommand{\su}{\hspace*{.1in}}
 
\newcommand{\etal}{$\rm et \; al.\,$}

\def\loe{\lower 0.6ex\hbox{${}\stackrel{<}{\sim}{}$}}
\def\goe{\lower 0.6ex\hbox{${}\stackrel{>}{\sim}{}$}}

\newcommand{\cen}[1]{\centerline{#1}}
 
\newcommand{\be}{\begin{equation}}
\newcommand{\en}{\end{equation}}
 
\newcommand{\cg}{2CG~135+1 }
\newcommand{\cgp}{2CG~135+1}
 
\newcommand{\lsi}{LSI~$+$61$^{\circ}$~303 }
\newcommand{\lsip}{LSI~$+$61$^{\circ}$~303}

\title{Monitoring the Gamma-Ray Source 2CG 135+1 and the Radio Star 
LSI +61$^{\circ}$ 303}

\author{M. Tavani\inst{1} 
\and W. Hermsen\inst{2}
\and R. van Dijk\inst{2,3}
\and M. Strickman\inst{4} 
\and S.N. Zhang\inst{5}
\and R.S. Foster\inst{6}
\and P.S. Ray\inst{6}
\and J. Mattox\inst{7}
\and M. Ulmer\inst{8}
\and W. Purcell\inst{8}
\and M. Coe\inst{9}
}
 
\begin{document}
\offprints{M. Tavani}
\institute{
{Columbia Astrophysics Laboratory, Columbia University, New
York, NY 10027, USA} 
% \\ e-mail: tavani@meleas.phys.columbia.edu }
\and
{SRON-Utrecht, Sorbonnelaan 2, 3584 CA Utrecht, The Netherlands}
\and
{Astronomical Institute, University of Amsterdam, Kruislaan 403, NL-1098 SJ
Amsterdam, The Netherlands}
\and
{Code 7651, Naval Research Laboratory, Washington, DC  20375, USA}
\and
{Universities Space Research Association, NASA/MSFC, Huntsville, AL, 35812, USA}
\and
{Code 7210, Naval Research Laboratory, Washington, DC  20375, USA}
\and
{Department of Astronomy, University of Maryland, College Park,  MD 20742, USA}
\and
{Department of Physics \& Astronomy, Northwestern University, Evanston, IL 60208, USA}
\and
{Department of Physics, University of Southampton, Southampton, SO17 1BJ, U.K.}
%\and
%{Jet Propulsion Laboratory, Caltech, Pasadena, CA 91109, USA} 
}

% \date{Received: October 20, 1995    ; Accepted    ,  }
 
\maketitle
 
\begin{abstract}

We report the results of 
%a phase 3 
a CGRO multi-instrument series of
observations  of the unidentified gamma-ray source 2CG 135+1 and of
its possible counterpart, the peculiar radio source GT 0236+610
coincident with the Be star  \lsip.
Previous 
%phase 1 and 2 EGRET 
observations of 2CG 135+1 are consistent
with a positional coincidence of the gamma-ray source with \lsip.
Since January 1994, we continuously  monitored the time 
variable radio source  GT 0236+610
with the Green Bank Interferometer.
For the first time, gamma-ray observations of 
2CG 135+1 can be correlated with an extensive database of radio 
observations of \lsip.  We present 
%preliminary 
OSSE and COMPTEL data obtained during
the  period May-June 1994, and BATSE data for the period
April 1994--January 1995.
%phase 3 observations,  
We  discuss the possible time variability of the \ggg-ray emission
and  spectral properties of 2CG~135+1.
% We also discuss a search for a possible correlation between
%the gamma-ray  emission from 2CG 135+1 and the radio emission of LSI 61 303.
%Additional data on the two sources are needed  to resolve the issue. 
%We discuss the theoretical implications of our results, and different
%A plausible model for the observed gamma-ray emission is
%shock emission
%models for the gamma-ray emission in a binary containing a compact star
%orbiting around a massive companion. 
Understanding the nature of 2CG 135+1
may be  of crucial importance for the interpretation of a class of unidentified 
and time variable gamma-ray sources in the Galaxy.
 

\end{abstract}

\normalsize

\section{Introduction}

%\vskip .1in
 
The source \cg is one of the most prominent unidentified
gamma-ray sources.
Since its discovery by  the  COS-B satellite
(Hermsen \etal 1977; Swanenburg \etal 1981),
 no satisfactory explanation has been found
for  the  nature of \cg and its gamma-ray emission mechanism.
The source is located near the Galactic plane
at Galactic coordinates $l = 135^{\circ}.74, b = 1^{\circ}.22$.
 
The COS-B error box (of approximately $1^{\circ}$ radius) most notably
contains  supernova remnants and OB associations, % (Montmerle, \etal  197...),
the low-redshift ($z=0.043$)  quasar QSO~0241+622,
% (Apparao \etal  1978),
and the radio source GT~0236+61 identified with the % B0I 
massive star
\lsi at the distance of $\sim 2.3$~kpc (Gregory \etal 1979).
The existence of \cg was %recently 
confirmed by  EGRET during CGRO phase 1, 2 and 3 observations
%{\it Compton Gamma-Ray Observatory} (
%CGRO observations carried out during the
%period 1991-1994. In particular, a relatively strong \ggg-ray
%source  was detected  by the 
%CGRO instrument 
%EGRET
%in the energy range 30~MeV-1~GeV during a CGRO phase-1 pointing in 1991
(von Montigny \etal  1993, Thompson \etal, 1995; Kniffen \etal, 1996).
The new  position of \cg  determined by EGRET during the GRO
phase~1-2 pointings (with an error box diameter of $\sim 20'$)
 is well within the old  COS-B error box 
 and is consistent with the position of \lsip.  % (see Fig. 1).
COMPTEL phase 1 and 2 observations confirm the existence of a source
%Also the CGRO instrument COMPTEL detected gamma-ray flux in the
in the energy range 1-30 MeV 
%from a source whose $\sim 1^{\circ}$ error box
%is consistent with the \lsi source 
(van Dijk \etal  1994, 1996).
It is 
%therefore  
possible that the 
%therefore natural to consider the possible association of the
\ggg-ray source \cg is associated
 with 
%ms the bright radio star
 \lsi (Taylor \& Gregory, 1982).
However, we emphasize that no unambiguous proof of a relation between the
%ms modulated 
radio source and the gamma-ray emission from \cg
 exists at the moment. This paper reports the most extensive
 multiwavelength campaign 
ever attempted to determine the nature of \cgp.

The radio source GT~0236+610 coincident with
the B0~Ve star LSI $+61^\circ$ 303
[at an estimated distance of
$d \simeq 2.3 $ kpc (Gregory \etal  1979;
Taylor \& Gregory, 1982)]
 is unique among the Be star systems
in its highly variable  and periodic radio emission.
%;
%Murdin et al. 1980).  %;  Haynes \etal  1980;
 %Taylor and Gregory, 1982).
The radio continuum shows a
 period { $P_{LSI} = 26.496 \pm 0.008$ days}
 (Taylor \& Gregory, 1984)
with a  
complex outburst behavior and a possible $\sim 4$ yr modulation
of outburst amplitudes and change of the radio
light curve pattern (Gregory \etal  1989).
Optical and UV observations of \lsi showed a spectrum typical of
an early-type Be star with P~Cygni profiles indicating mass
loss through an equatorial disk
with surface temperature $T \sim 3 \cdot 10^4$ K
and 
 bolometric
 %ms star 
luminosity $L_B \sim 10^{38}$ erg/s
%(Hutchings and Crampton, 1981; Maraschi, Tanzi \& Treves, 1981;
% (Paredes \& Figeras, 1986; 
(Paredes \etal  1994).
Soft X-ray emission in the range 0.5--4 keV
from \lsi has been detected by {\it Einstein} with a luminosity
$L_X \sim 10^{33} \ergs$ (Bignami \etal  1981).
%Share \etal 1979; Bignami \etal  1981).
Recent observations of \lsi by { ASCA} (Harrison \etal  1996)
and ROSAT (Goldoni \& Mereghetti, 1995)
confirm the old detection by {\it Einstein}.
%
% No clear evidence of modulated  soft X-ray emission has as yet been 
% obtained.
%
%observation of \lsi in the energy range 0-5-10~keV
%(Harrison \etal  1994) confirms the old { Einstein} detection
%of \lsi.
%We also note that
%{ ROSAT} observations of \lsi were recently carried
%out, with an interesting positive detection (Goldoni \& Mereghetti, 1995).

%\newpage
%\vskip .3in
\section{BATSE Data Analysis}
%\vskip .1in

\vskip .1in
\begin{table*}
{\normalsize
\begin{center}
\cen{\sl Table 1: BATSE data on 2CG 135+1}
\vskip .05in
\begin{tabular}{lccc}
\hline
TJD  & Dates & average flux (mCrab) &
Flux (mCrab) near $\phi_{LSI} = 0.8$\\
\hline
8373-8823    & Apr.27,1991-Jul.20,1992 & $3.4 \pm 1.0$ &  $13.6 \pm 2.8$\\
8834-9289    & Jul.31,1992-Oct.29,1993 & $1.3 \pm 1.2$ &  $11.2 \pm 3.2$\\
9290-9748    & Oct.30,1993-Jan.31,1995 & $2.8 \pm 1.0$ &  --- \\
\hline
\end{tabular}
\end{center} }
\end{table*}

\begin{figure}[thbp]
 % \picplace{4cm}
%\centering{ 
%\vspace*{-.5cm}
%\psfig{figure=135.batse.ps,height=8.cm,width=8.8cm} }
\psfig{figure=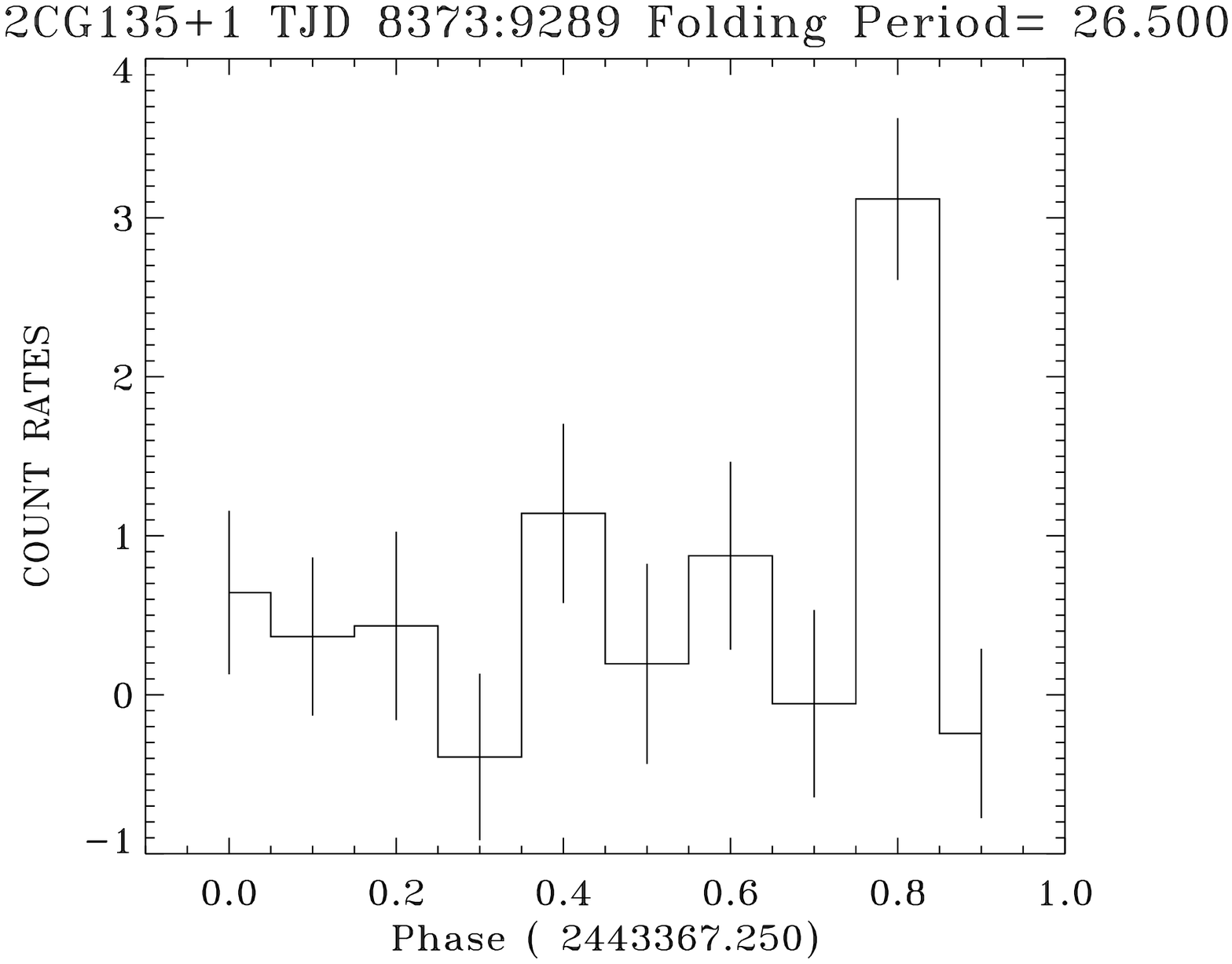,%This percentage sign is vital here
bbllx=50bp,bblly=72bp,bburx=568bp,bbury=478bp%So is this one
,width=3.5in,clip=}  %}
 \caption{
BATSE folded light curve ($P=26.496^d$) of \cg for the period
(truncated Julian days) TJD 8373-9289 (April 27, 1991 - October 29, 1993).
1 count/s corresponds to 3mCrab and errors are 1-$\sigma$.
Data from Zhang  \etal (1996a).}
 \label{Figure1}
\end{figure}

We analysed BATSE data (large area detectors, LADs)
for a $\sim 0.^{\circ}5$ radius region 
centered on the position of
2CG~135+1. BATSE data were studied for a period starting in April~1991
until  January 1995
% the end of 1994 
and processed by the Earth
occultation analysis technique (Harmon \etal  1993).
  % of the large area detectors have been analysed since the
% beginning of the CGRO mission in April 1991 until very recently. The Earth
%occultation analysis technique is applied to this data analysis. The
No emission in the range 20--200~keV at the level of
$\sim 50$~mCrab or more was detected from the direction of \cg for
standard integration
times of order of $\sim$~1-2~weeks.
  However, using the entire 3.5 years of
  BATSE data yields a possible detection with an average flux of 
  $2.4 \pm 0.6$ mCrab.
%However, extending in time the BATSE data   baseline
% %\footnote{
% %The Earth occultation technique allows to extend the data baseline
% %for long durations.}
  % leads to a possible
% detection of emission whose average flux from all BATSE data 
% is about $2.4 \pm 0.6$~mCrab. 
This level of emission is consistent
with COMPTEL and OSSE low-energy  extrapolations of the \cg flux,
even though other sources might contribute.
%Preliminary results indicate a possible detection of this source. The
%average flux from all BATSE data is about 2.4 +/- 0.6 mCrab, compatible with
%the power law extrapolation from OSSE, COMPTEL and EGRET measurements at higher
%energies. 
In order to test a  possible association of the emission detected by 
BATSE with \lsip,
 we performed an epoch-folding analysis of the BATSE data  for different
modulation periods.
For practical purposes, the BATSE data from April 27 1991 to January 31
1995 were divided into three equal
 time intervals.
A peak of the BATSE  folded light curve appears at the
radio phase $\phi \sim 0.8 \pm 0.05$ only for folding periods in
the interval $P_{B} = 26.5  \pm 0.05$.
This period range includes $P_{LSI}$
and it is possible that the emission near the BATSE peak originates from
\lsip.
 Fig.~\ref{Figure1} and 
Table~1 give the  details of the BATSE  results (Zhang \etal  1996a).

%\begin{figure}
 %\picplace{9cm}
 
%\vspace*{-1.5cm}

 Without any {\it
a priori} knowledge of the peak occurrence in phase, we find a total
%the total chisquare is
$\chi^2 = 33.6968 $
for 9 degrees of freedom, implying a confidence level of 0.9999 against the
(null) hypothesis of a flat lightcurve.
The  channel centered near phase $\phi= 0.8$
 alone contributes 30.8024 to the total $\chi^2$, meaning
the probability that this peak is a result of statistical fluctuation
is less than 
$ 3.2 \times 10^{-4}$.
If {\it a priori } knowledge is available about the peak location near phase
$\phi= 0.8$, the significance of this peak is significantly   higher
(probability less than 2.5$\times$10$^{-7}$ of a statistical fluctuation). 
We have searched for a periodicity from 26.0 days
to 27.0 days with a step-size of 0.1 day. No statistically significant peak
was found at a period other than  near 26.5 days.
The obvious systematic effect due to
the satellite orbital precession   cannot produce a periodic signal
of period near 26.5 days. 
The combined effect of airdrag and the CGRO  orbit
reboost have resulted in the satellite orbital precession period varying from
$49.5 \pm 0.5$~days 
to $52.5 \pm 0.5$~days. Therefore  a sharp peak for
a folded light curve with a period $\sim 26.496^d$   is
unlikely caused by  the CGRO satellite orbital precession.
We note that the same folding technique has also
been applied to other sources, resulting  in the 
BATSE detection of  binary eclipses
(1.4 days) and accretion disk precession (30.5 days) of LMC X-4 (Zhang
\etal 1996b)
and the discovery of a 241~day period
transient source GRO J1849-03 (Zhang \etal  1996c).  % these processings).
Nevertheless, we believe that the existence of hard X-ray emission
related to  the activity of the radio source GT~0236+610  needs to
be confirmed by additional data.
We are currently carrying out  
a thorough  examination of the overall statistical 
and systematic significance, in particular
studying all possible sources of systematic effects.

The BATSE  peak emission appears only during the first two time intervals
from mid-1991 to most of 1993.
%A peak at the radio phase 0.8 is
%This 
% The appears peak is detected for the data of the 
% first two sections at flux levels of 13.6 +/- 2.8 and
% 11.2 +/- 3.2 mCrab respectively. The peak is, however, absent from the data of
% the third section. The average flux levels are 3.4 +/- 1.0, 1.3 +/- 1.2 and 2.8
% +/- 1.0 mCrab for the first, second and third sections respectively. 
We note that drastic changes of the radio light curve of \lsi  from the average
were reported in the literature, with radio peaks observed near phase $\sim 0.8$
(Gregory \etal  1989; Paredes \etal 1990; Taylor \etal 1995).
 An interesting  systematic shift as a function of time of phases
 corresponding to  peak radio maxima 
of GT~0236+610  is evident since the beginning of 1994
in  the 2.25 and 8.3~GHz data obtained at Green Bank (Fig.~\ref{Figure4}).
BATSE data provide support for  time variability of the hard X-ray emission
from the direction of 2CG~135+1 within a timescale of several years, and future data
will be important in proving a possible long-timescale trend.

% 8373-8823    & Apr.27,1991-Jul.20,1992 & $3.4 \pm 1.0$ &  $13.6 \pm 2.8$\\
% 8834-9289    & Jul.31,1992-Oct.29,1993 & $1.3 \pm 1.2$ &  $11.2 \pm 3.2$\\
% 9290-9748    & Oct.30,1993-Jan.31,1995 & $2.8 \pm 1.0$ &  --- \\
% \hline
% \end{tabular}
% \end{center}

%\vskip .3in

\section{  OSSE Data Analysis}
%\vskip .1in
 
\begin{figure}[thbp]
 % \picplace{4cm}
\centering{ \vspace*{-.5cm}
\psfig{figure=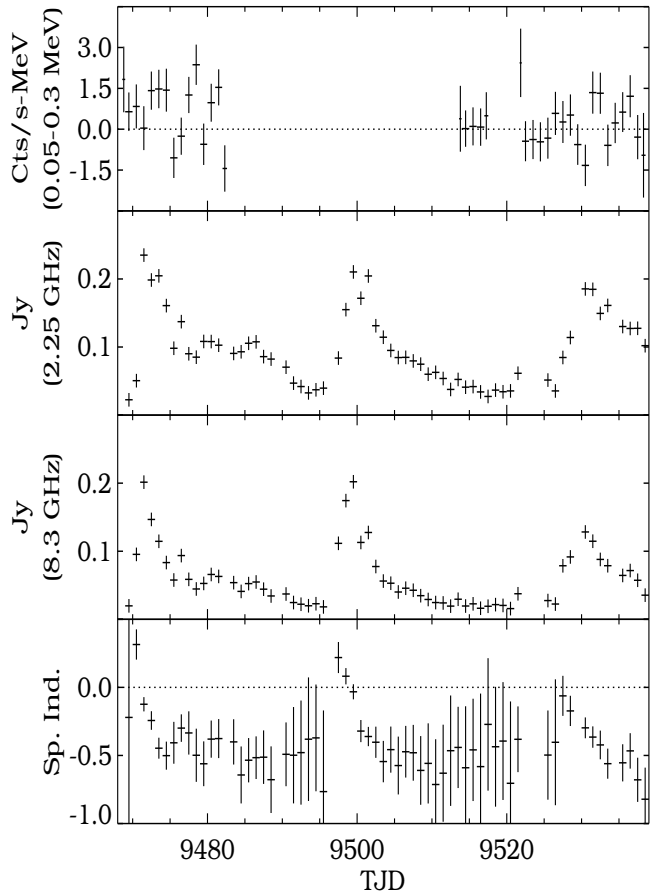,height=13.cm,width=8.8cm} }
 \caption{OSSE light curve of 2CG~135+1
and Green Bank radio light curve (at frequencies 2.25 and 8.3~GHz)
and radio spectral index $\delta$ (defined as $S_{\nu} \propto \nu^{\delta}$)
of GT~0236+610  for the three GRO 
phase 3 observations in May-June 1994
(from Strickman \etal  1996 and Ray  \etal, 1995).}
 \label{Figure2}
\end{figure}

\begin{figure}[thbp]
 % \picplace{4cm}
\centering{ 
%\vspace*{-.5cm}
\psfig{figure=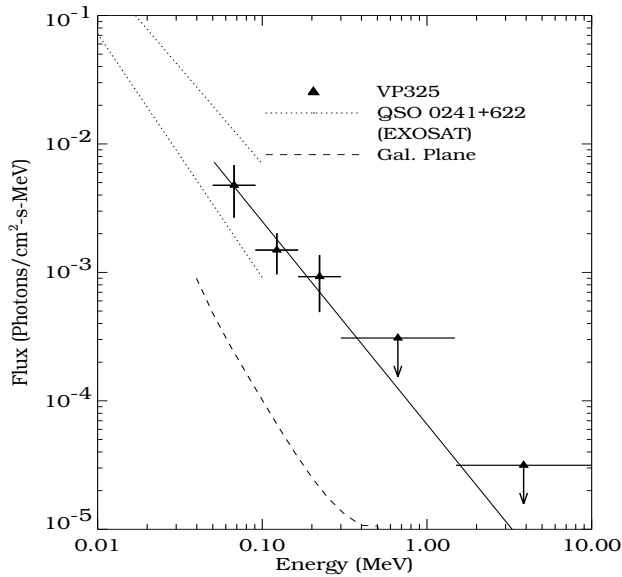,height=8.cm,width=8.5cm} }
 \caption{OSSE  spectrum of \cg observed during VP 325.
The extrapolated spectrum of  QSO~0241+622 and the level of diffuse
Galactic bakground are indicated
(data from Strickman \etal  1996.}
 \label{Figure3}
\end{figure}

OSSE observed the region containing \lsi 
 on three occasions from April
through July of 1994, as shown in Table~2.  The table indicates the dates of
the observations, the total OSSE on-source time in detector-seconds (i.e. the sum of
on-source  times for all OSSE detectors used in the observation) and the
estimated phases of  the radio outburst cycle.
%, using the ephemeris of Taylor \& Gregory (1984).
 %  Note that the minimum detectable fluxes for OSSE as indicated
% in Johnson et al. (1993) are based on a $1 \times 10^{6}$ detector-second
% observation, hence the instrument sensitivity for these observations are
% reduced from those quoted by factors of 0.6, 0.3 and 0.6 respectively for the
% three observing periods.
Fig.~\ref{Figure2} shows
  the OSSE light curves for the three observing periods
together with simultaneous radio GBI data.
 Fig.~\ref{Figure3} gives the OSSE  spectrum for the GRO viewing period
(VP)~325.

%\begin{table}[h]
\begin{center}
{\sl  Table 2: OSSE observations of \cg}
\vskip .05in
\begin{tabular}{lllc}
\hline
VP &  Dates & $\Delta \phi_{LSI}$ 
%\lsi phase interval
 & Exp.  ($10^{5}$~s)\\
%Det-seconds)\\
\hline
325 &  26 Apr-10 May 94 & 0.31$-$0.84 & 3.3\\
330 &  10 Jun-14 Jun 94 & 0.01$-$0.16 & 0.98\\
332 &  18 Jun- 5 Jul 94 & 0.32$-$0.95 & 3.6\\
\hline
\end{tabular}
\end{center}
%\end{table}

The OSSE field of view is large enough that nearby sources may  contribute to
the detected emission.
In particular, during all three viewing periods, the nearby source
QSO~0241+622 was  observed with significant exposure.  QSO~0241+622 is a
low-luminosity quasar that is well known to be an X-ray emitter.  Turner \&
Pounds (1989) have reported that EXOSAT observed this object with a spectrum
represented by a power law with photon  index 1.7 up to $\sim 20$~keV. 
% Bassani et al. (1992)
%report on a SIGMA observation of QSO~0241+622 in which, while they achieved
%only upper limits, they could not rule out emission consistent with the
%extrapolation of the EXOSAT result to higher energies.  
% Figure (2) shows the
% extrapolation to OSSE energies of the EXOSAT spectrum (within the indicated
% error band), corrected for the fact
% that QSO~0241+622 was at 70\% response in the OSSE collimator.
  The OSSE
results are consistent with the extrapolation, hence the possibility that the
emission observed by OSSE is from QSO~0241+622 cannot be ruled out
conclusively by spectral analysis alone.
% 
%\begin{figure}
 %\picplace{9cm}
OSSE detected a statistically significant flux ($\sim 4\sigma$)
 from the direction of \cg in the 50 - 300 keV band
only during the GRO  VP~325.
%Of the three observations, only during the first (labeled viewing period 325
%in the table) did OSSE detect a positive flux from the field of view
%containing LSI 61 303. 
 %In the 50 - 300 keV band, flux is detected at a
%statistical significance $\sim 4\sigma$. 
 The spectrum, shown in Fig.~\ref{Figure3} 
is well-represented by a power-law model with index $1.6^{+0.6}_{-0.5}$.
%  The
%spectra measured for VPs 330 and 332 also are shown in 
%Fig.~(\ref{Figure2}); neither 
OSSE did not detect \cg during VPs 330 and 332
 at better than $2.5\sigma$. 

%As shown, there is no
Given the relatively low statistical significance of the detection,
%during VP 325, 
there is no significant evidence of
time variability in OSSE data
%evidence for variability
 from one viewing period to the next.
% 
%Figure ???+1 shows the flux in 50-300 keV band for the three viewing periods
%as a function of time with one-day resolution. 
However, we also note that the OSSE data for VP~325
% the only viewing
%period during which a significant flux is detected, is inconsistent with a
are inconsistent with a 
constant flux at the 98.9\% ($\sim2.5\sigma$) level.
A comparison between OSSE and simultaneous radio data of \lsi
(cf.,  Fig.~\ref{Figure2})  does not
support a correlated behavior at a statistically significant level.

%\begin{table}[h]
% \begin{center}
% {\bf Table 1: OSSE observations of the \cg region}
% \begin{tabular}{|c|l|l|l|c|}
% \hline
% GRO VP & TJD    & Dates & $\Delta \phi_{LSI}$
% %\lsi phase interval
 % & Exposure ($10^{5}$~s)\\
% %Det-seconds)\\
% \hline
% 325 & 9468.61 - 9482.61 & 26 Apr-10 May 94 & 0.31 - 0.84 & 3.3\\
% 330 & 9513.63 - 9517.53 & 10 Jun-14 Jun 94 & 0.01 - 0.16 & 0.98\\
% 332 & 9521.72 - 9538.55 & 18 Jun- 5 Jul 94 & 0.32 - 0.95 & 3.6\\
% \hline
% \end{tabular}
% \end{center}
% %\end{table}
 
%\begin{figure}
% \picplace{5cm}

%\vskip .3in
\section{ COMPTEL Data Analysis}
%\vskip .1in

% COMPTEL observed the 2CG135+1 region from April 26 up to May 10 1994
% (Obs. 325). The corresponding radio-phase interval for this observation
% is 0.31-0.84, which contains the phase of the peak radio outbursts
% of the proposed counterpart GT 0236+610.
 
COMPTEL data for VP~325 were analysed in the 4 standard energy intervals 0.75-1 MeV,
1-3 MeV, 3-10 MeV and 10-30 MeV, using a maximum likelihood method to
obtain source significances, fluxes and $1\sigma$ errors. 
%Hereby c
Corrections
for deadtime and energy-dependent selection effects have been taken into
account. Emission from the Crab nebula (which contributes as many
events to the dataspace as 2CG~135+1 in this observation)
was included in the data analysis.
%In the analysis, a model for the Crab, which contributes as many
%events to the dataspace as 2CG135+1 in this observation, had to be included as
%well. 
The quoted fluxes still need to be corrected for the possible
contribution of diffuse Galactic emission in this region.

Using the complete set of data for VP~325, we obtain a detection
significance of $2.5\sigma$ at the position of 2CG135+1 in the 3-10 MeV energy
range.
% Fig. 8  shows the $1\sigma$ and $2\sigma$ error location contours for this
% excess, with 2CG135+1 within the $2\sigma$ contour.
For the other energy ranges, only upper limits are obtained. 
%Table~3 gives  the fluxes for each of the energy ranges.
Due to the low signal-to-noise of the VP~325 detection, 
no correlation of COMPTEL and radio flux from \lsi can be established.
%nothing can be said about any possible
%correlation of the 3-10 MeV flux with the radio phase of GT 0236+610,
%the proposed counterpart.
 We note that the VP~325 COMPTEL flux 
% Note, however, that the flux
is consistent with the time-averaged flux of VP~15+31+34
(van Dijk et al. 1994, 1996).

\def\cmsmev{\hbox{cm}^{-2}\hbox{ s}^{-1}\hbox{ MeV}^{-1}}

%\vskip .2in

% \begin{center}
% \centerline{\bf Table 3: COMPTEL flux from \cg during VP~325 }
% %  Upper limits}
% % \vskip .08in
% \begin{tabular}{|c|c|} 
% \hline
        % Energy channel (MeV)   &  Flux or $2\sigma$ upper limit ($\cmsmev$)\\
% \hline
        % 0.75-1   &        $<4.9\times 10^{-4}$\\
        % 1-3      &       $<1.1\times 10^{-4}$\\
        % 3-10     &       $(9.0\pm3.7)\times 10^{-6}$\\
        % 10-30    &       $<1.5\times 10^{-6}$\\
% \hline
% \end{tabular}

% \end{center}

%\vskip .3in
\section{ Radio Monitoring of \lsi at Green Bank}
%\vskip .1in

%The NRL-GBI Monitoring Program currently emphasizes monitoring of X-ray
% binaries on a daily basis.  The Green Bank Interferometer is operated by the
% National Radio Astronomy Observatory for the Naval Research Laboratory and is
% funded by the Office of Naval Research and multi-frequency guest observer 
% grants from NASA.  
% A limited amount of guest observing time is available, and 
% outside investigators are encouraged to either obtain grants for support for 
% the GBI or to collaborate with NRL staff scientists. 
 
\begin{figure}[thbp]
 % \picplace{4cm}
\centering{
\vspace*{-.7cm}
\psfig{figure=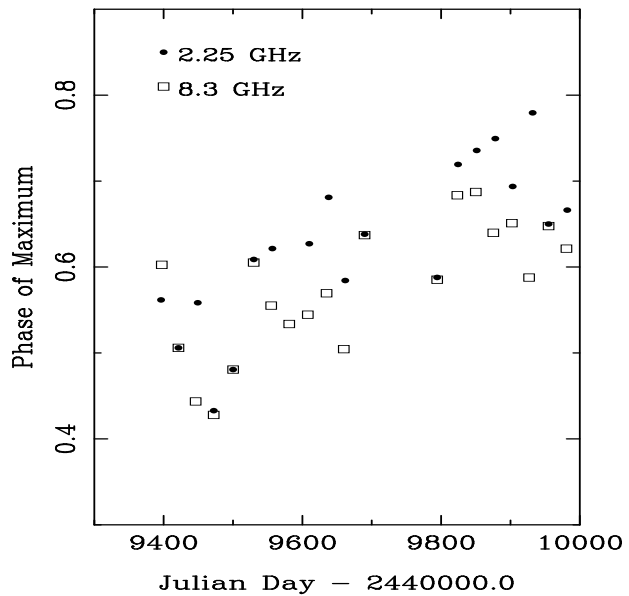,height=9.cm,width=9.5cm} }
 \caption{Time behavior of the radio peak phases  of GT~0236+610 
as continuously monitored at 2.25 and 8.3 GHz at the Green Bank
interferometer since the beginning of 1994.
A folding period of $P=26.496^d$ is assumed (data from Ray \etal  1995).}
 \label{Figure4}
\end{figure}

     The Green Bank Interferometer 
(GBI)  consists of two 85~ft antennas on a 2.4 km baseline and
observations are simultaneously carried out at two frequencies (2.25 and 8.3 
GHz).
%, and they  may be scheduled within 5 hours of the local meridian. 
 Calibration
is usually referred to four standard calibration sources.  
% Because activity in 
% radio sources develops over 
% a longer time period than the corresponding X-ray 
% or gamma-ray counterpart, simultanous multi-frequency observations alone are 
% not appropriate, and radio monitoring for an extended time period is 
% encouraged. 
We started in January 1994
 an extensive program of  radio monitoring of GT~0236+610.
Fig.~\ref{Figure2}
 gives the radio  light curve of GT~0236+610 within the time period
 of interest here.
%showing interesting cycle-to-cycle variations of the emission pattern.
During most of 1994
the amplitude of the outbursts is typically in the range 200-300 mJy,
and the phase range of the peak emission is $\Delta_{LSI} \sim 0.4-0.6$.
The radio peak amplitudes and phases observed at GBI during 1994 
are in agreement with those previously detected near the maximum
of a  possible  4-year modulation cycle (Gregory \etal  1989).
%However, it is possible that the peak phase range is subject to
%a systematic shift as a function of time.
%Fig.~(\ref{Figure4}) shows the  behavior of the radio peak phase
%as a function of time as continuously monitored at GBI for the first
%time during more than a year and one half period.
However, we notice an interesting
 systematic trend  of the peak emission phase
as time progresses with a shift from phase $\phi \sim 0.5$ near 
truncated Julian day (TJD) 9400
 to $\phi \sim 0.7-0.8$ near TJD~9800-9900
of the $26.496^d$ cycle. % Fig.~(\ref{Figure4}).
The  radio outburst
peak phases of  GT~0236+610 observed at 2.25 and 8.3~GHz by the GBI 
 are shown in Fig.~\ref{Figure4}.  
%These phases are measured by cross-correlating each outburst with a standard template.  These measurements are more robust against
%missing data or interference peaks than just selecting the highest
%single measurement as the peak.
% 
%
This effect is detected for the first time in the GT~0236+610 source,
and it will be important to keep monitoring the radio emission
to confirm the reality of a multi-year systematic behavior of the
radio flare timing. 
This
issue is of the  greatest importance
given the possible relation of the
time behavior of the radio peak phase
with the hard X-ray enhancement detected by BATSE.
Sporadic radio observations of GT~0236+610 detected peak maxima
at phases near 0.8-0.9 in the past (Gregory \etal  1989;
Paredes \etal  1990; Taylor \etal, 1995). However, no evidence
of a systematic trend was obtained before the GBI monitoring.
In principle, the systematic peak phase shift 
 might be due to either a real phase change
or to a change of orbital period.
It is also possible that the systematic peak phase change is related 
to the $\sim 4$-year modulation of the peak amplitude previously reported
by Gregory et~al. (1989).
 
% \picplace{4cm}
 
% \begin{figure}[thbp]
 % % \picplace{4cm}
% \centering{ \vspace*{-.5cm}
% \psfig{figure=lsi.radio.gbi.symp.ps,height=4.cm,width=9.cm} }
% \caption{Radio light curve (2.2 GHz) 
%  of GT~0236+610 as monitored by the
% Green Bank interferometer (data from Foster \etal  1995)} 
% \label{Figure3}
% \end{figure}

\vspace*{-.05cm} 
From the time variability of the radio emission  of GT~0236+610, 
it is clear that only  a long-term
radio monitoring  simultaneous  with \ggg-ray observations  can
ensure a meaningful comparison of radio and \ggg-ray data.

\begin{figure*}[thbp]
% \picplace{8cm}
\centering{
\vspace*{-0.5cm}
\psfig{figure=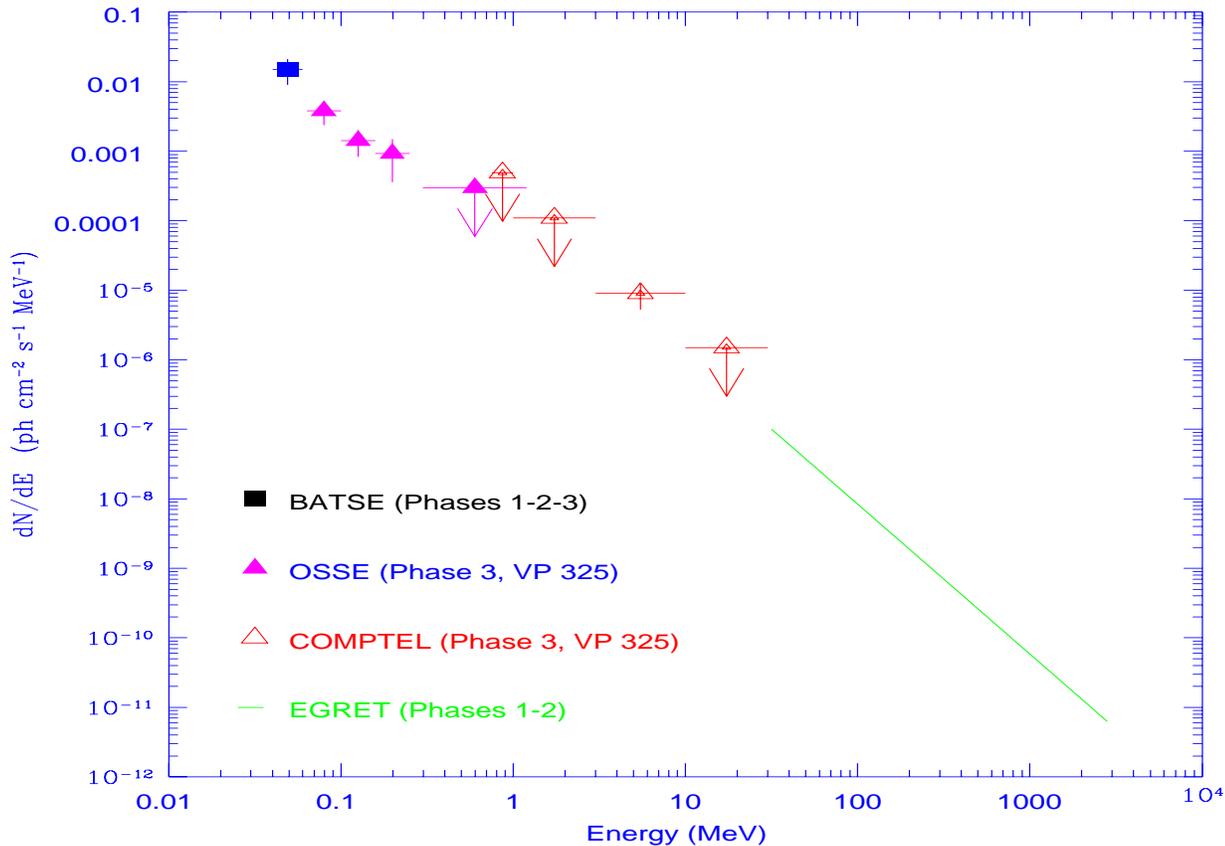,height=12.cm,width=17.cm} }
 \caption{CGRO multi-instrument   spectrum of \cgp: possibly
simultaneous spectral components are 
given for a broad-band energy interval.}
 \label{Figure5}
\end{figure*}

% \newpage
% \vskip .3in

\section{ Discussion}

%\vskip .1in
Multi-instrument CGRO observations provided crucial information on
the  the enigmatic \ggg-ray source
\cgp. No unambiguous  relation between the high-energy and radio emission
from \cg and its possible counterpart GT~0236+610 could be
established. 
It is possible that the EGRET source is not related with the \lsi system
and only future correlated EGRET and multiwavelength observations can
establish a real connection. 
However, several interesting aspects of the emission from \cg were
revealed and future observations will be able to confirm the interesting
hints of a  correlated behavior. The 
VP 325  observations confirm the
existence of a relatively weak
source in the 3-10~MeV range consistent with the position
of \cgp. At lower energies, the  low flux prevents an unambiguous
detection of \cg by OSSE and BATSE for  integration periods of order
of 1-2 weeks.  Contamination from the source QSO~0241+622 cannot be in
principle excluded in OSSE data.
We note that in the 0.1-10~MeV energy range, the 
OSSE/COMPTEL upper limits (or weak detection during VP~325) are about one
order of magnitude less than the flux reported by Perotti \etal~(1980).
%from the same region
%by the MISO instrument in 1978 (Perotti \etal  1980).
Emission below $\sim$~1~MeV appears to be variable within a timescale
of a few weeks.
% (cf., Fig.~\ref{Figure2}).
It is clear that the hard X-ray/soft \ggg-ray emission of \cg is not
strongly correlated with the radio flares of GT~0236+610
(cf., Fig.~\ref{Figure2}). The OSSE detection of
 a weak hard X-ray flux 
during the period April 26-May 10, 1994 is coincident
with the onset and decay
of a relatively strong radio outburst of GT~0236+610.
No flux was detected by OSSE during VP 330 (coincident with a 
shallow minimum of the radio emission from GT~0236+610) and
the weakness of the source prevents to establish
 an unambiguous correlation between hard X-ray and radio emission
for VP~332 (see Fig.~\ref{Figure2}). 
The possible relation between the hard X-ray emission  detected by BATSE
and the radio long-term systematic change needs to be confirmed by
more data. We cannot even exclude that the BATSE source is not related
to the EGRET source.  Only future data showing
a  clear modulation of the \ggg-ray flux with the \lsi period
and a confirmation of the relation between BATSE and radio data
will establish this connection.

Fig.~\ref{Figure5} shows the 
%current status of a 
combined multi-instrument  spectrum of
emission from the \cg region for VP~325; the 
%preliminary
 OSSE and COMPTEL data
are plotted with the non-simultaneous average EGRET spectral flux
and  with the BATSE low-energy average flux.
% Our current BATSE data cannot establish unambigously 
% the possible existence of a  long timescale modulation ($\sim$
% years) of the hard X-ray emission.
%
%
%
% The possible BATSE detection during the 1991-1993 period
 % of modulated flux at the level of
% a  few mCrab in the 20--200 keV
% range  deserves a  careful analysis.
% Based on the radio  light curve of Fig.~(9),
% the existence of peaked emission near $\phi_{LSI} \sim 0.8$
% would suggest that the  emission is
% not correlated with the radio outbursts [typically centered
% around $\phi_{LSI} \sim 0.5-0.6$ for the data of Fig.~(9)]. 
% However, we notice that drastic
% pattern   changes of the radio light curve of GT~0236+610
% (possibly related to the 4-year radio peak modulation cycle)
% were reported, with outburst peak observed near $\phi_{LSI} \sim
% 0.8$ (Gregory \etal  1989).
% % Future  radio and \ggg-ray data are therefore crucial to resolve the 
% % issue of the possible association of \cg with \lsip.
Future continuous radio and   BATSE coverage of the 
emission from \cg  together with new pointed EGRET,  COMPTEL 
and OSSE observations  scheduled in 1996
will contribute to confirm or disproof a   correlation between
the radio and \ggg-ray emission from the \cgp/\lsi complex.

% \subsection{Theoretical considerations}

Two main models for the radio and high energy emission from
\lsi are currently under debate.
 One model suggests that the radio outbursts
  (possibly correlated with the high energy
emission) are  produced by
streams of relativistic particles originating in
episodes of super-Eddington {\it accretion}
onto a compact star embedded in the mass outflow of the
Be star companion (e.g., Taylor \etal  1992).
No \ggg-ray emission with photon energy larger than
10~MeV has ever been observed from an accreting source,
and the nature of the  accretion process producing a spectrum
of the kind of Fig.~\ref{Figure5} must be different from
other known sources.

Alternately,
\lsi might contain  a  {\it non-accreting  young pulsar} in  orbit
around the  mass-losing Be star (e.g., Maraschi \& Treves, 1981;
Tavani, 1995).
In this case, the high-energy emission can provide
  an important diagnostics
of the shock emission region where the pulsar wind interacts with
the circumstellar material originating from the surface of the
massive Be star (Tavani, 1994).
The modulation of the radio emission might be due to the time variable
geometry of a `pulsar cavity' as a function of the orbital phase,
and a systematic phase shift of the peak radio emission can  be
a consequence of quasi-cyclic pulsar/outflow interaction for a
precessing Be star outer disk.
High-energy (X-ray and \ggg-ray) emission can be produced by relativistic
shocked pairs of the pulsar wind by synchrotron and inverse Compton
emission, a  mechanism observed in the case of the Crab
nebula (e.g., Kennel \& Coroniti, 1984) 
and which has been recently shown to operate
in the Be star/pulsar PSR~1259-63 system  (Tavani \& Arons, 1997,
hereafter TA97).
For an observed
 efficiency of conversion of spindown pulsar energy into
hard X-ray/\ggg-ray
 shock emission between 1\% and 10\% (Kennel \& Coroniti, 1984;
TA97),
we deduce a pulsar spindown luminosity 
larger than  $\sim 10^{35}-10^{36} \,
\rm erg \, s^{-1}$ for a source at 2~kpc. 
The high-energy shock emissivity depends on the geometrical
and radiative characteristics of the pulsar cavity as the pulsar
(whose pulsed radio  emission is presumably `hidden' and  absorbed in the
companion outflow) orbits around the massive star \lsip. 
It is worthwhile to notice that a 
time variable mass outflow from the companion star
can produce a variable size of the pulsar cavity and of its high-energy
emission. Long  timescale modulations of the radio and
high-energy emission from a pulsar cavity can in principle be produced
(as clearly shown in the PSR~1259-63 system, cf. TA97). 

Cyclic changes in the geometry of the pulsar/outflow interaction 
(possibly due to precessional motion of the Be star outer disk) may
cause the changing pattern of the radio light curve of GT~0236+610.
Future  multiwavelength observations will be crucial.
Understanding the emission mechanism of sources such as \cg can
greatly help the interpretation of Galactic time variable \ggg-ray sources
%recently 
detected by CGRO.

\acknowledgements{
We thank Elizabeth Waltman for her invaluable support in processing
GBI data of GT~0236+610. Research supported by NASA grants NAG~5-2729 (MT)
and NAG5-2833 (JM).}

\normalsize

\section{References}

\def\nn{\noindent}
 
\noindent
Bignami, G.F.  \etal,  1981, ApJ, 247, L85.
 
% \noindent
% Clear, J., \etal,  1987, A\&A, 174, 85.
 
%\noindent
%Coe, M.J., Quenby, J.J. \& Engel, A.R., 1978, Nature, 274, 343.

%\nn 
%{Foster, R., etal., 1995, in preparation}
 
\nn
{Goldoni, P. \& Mereghetti, S., 1995, A\&A, in press}
 
\noindent
Gregory, P.C., \etal,  1979, AJ, 84 1030.

%\rref{Gregory, P.C. \& Taylor, A.R., 1986, ApJ, 92, 371}
 
\noindent
\rref{Gregory, P.C., \etal, 1989,
% Xu, H.J., Backhouse, C.J. \& Reid, A., 1989,  
ApJ, 229, 1054}

\rref{Harrison, F.A., 
%Leahy, D.A., Waltman, E.B., 
et al., 1996,  submitted to ApJ}
%in preparation.
 
% \noindent
 \rref{Harmon, B.A., \etal, 1994,  in 
%Proc. of the 2nd {\it Compton Observatory Symposium}, 
AIP Conf. 
%Series 
no. 304, p. 210}
 
\noindent
Hermsen \etal,  1977, Nature, 269, 494.

% \noindent
% Hermsen, W. \etal   1993, A\&A Suppl. Ser., 97, 97.
 
%\noindent
%Haynes, R. F., Lerche, I., and Murdin, P. 1980, A\&A, 87, 299.

\nn
Kniffen, D., \etal,  1996, submitted to ApJ.
 
\nn
Kennel, C.F. \& Coroniti, F.V.,  1984, ApJ, 283, 694.
 
\noindent
Maraschi, L., and Treves, A., 1981, M.N.R.A.S.  194, 1P.
 
%\noindent
%Maraschi, L., Tanzi, E.G. \& Treves, A., 1981, ApJ,  248, 1010.
 
%\noindent
%Monaghan, J.J., 1992, Ann. Rev. Astron. \& Astrophys., 30, 543.
 
% \noindent
% Murdin, P, \etal 
% Jauncey, D. L., Haynes, R. F., Lerche, I., Nichilson, G. D.,
  %Holt, S. S., and Kaluzienski, L. J.

%\noindent
%Paredes, J.M. \& Figueras, F., 1986, A\&A, 154, L30.

\nn
{Paredes, J.M., \etal,  1994, A\&A, 232, 377}
 
\nn
{Paredes, J.M., \etal,   1994, A\&A, 288, 519}
 
\nn
{Perotti, F., etal., 1980,  ApJ,  239, L49}

%\noindent
% \rref{Share, G.H. \etal  1979, in Proc. 21st COSPAR Meeting,
% (Innsbruck), eds. W.A. Bally \& L.E. Peterson, (New York: Academic),
% p. 535}
% 

\nn
\rref{Ray P.S.,  Foster R.S.,  Waltman E.B., Ghigo F.D.
%Johnston K.J. \&
\& Tavani, M., 1996, submitted to ApJ}

\nn
\rref{Strickman, M., 
%Tavani, M., Coe, M. 
\etal,  1996, in preparation}
 
\noindent
Swanenburg, B.N., \etal,  1981, ApJ, 243, L69.
 
\noindent
\rref{Tavani, M., 1995,
in %the Proceedings of the Winter School Conference
{\it ``The Gamma-Ray Sky with GRO and SIGMA},
%Les Houches (France), January 25-February 4, 1994,
eds. M. Signore, P. Salati \& G. Vedrenne,
(Dordrecht: Kluwer), p. 181}
 
\nn
Tavani, M. \& Arons, J., 1997,  ApJ, in press (TA97).
 
\noindent
Taylor, A. R. and Gregory, P. C. 1982, ApJ, 255, 210.
 
\noindent
Taylor, A. R. \&  Gregory, P. C. 1984, ApJ, 283, 273.
 
\noindent
\rref{Taylor, A.R., 
%Kenny, H.T., Spencer, R.E. \& Tzioumis, A., 
\etal, 1992, ApJ, 395, 268}

\rref{Taylor, A.R. \etal,  1996, A\&A, 305, 817}

\noindent
\rref{Thompson, D., \etal,  1995, 
%2nd  EGRET Catalog, 
ApJS, 101, 259}

\noindent
Turner, T.J. \& Pounds, K.A., 1989, MNRAS, 240, 833.
 
\noindent
\rref{van Dijk, R., \etal,  1994, AIP Conf. Proc. no. 304, p. 324}

\noindent
van Dijk, R., \etal, 1996, A\&A, in press.

%paper in preparation}
% in the Proceedings of the {\sl Second Compton Gamma-Ray Observatory Symposium},
% St. Louis, September 20-22 1993, eds.  J.P. Norris, N. Gehrels,
% C. Fichtel, AIP Conference Proceedings, in press}
%  

\rref{von Montigny C., \etal , 1993, {IAU Circ.} no. 5708}  
%(EGRET Collaboration), {IAU Circ.} no. 5708, Feb. 13, 1993}
 
%\noindent
%Waters, L.B.F.M., \etal  1991, A\&A,  244, 120.

\rref{Zhang, S.N., et al,  1996a, in preparation}

\rref{Zhang, S.N.,  et al, 1996b, these Proceedings}

\rref{Zhang, S.N.,  et al, 1996c, these Proceedings}

\end{document}